\documentclass[pre,reprint,nofootinbib,notitlepage,a4paper]{revtex4-1}

\usepackage{amsmath}
\usepackage{amssymb}
\usepackage{epsfig}
\usepackage{color}

\newcommand{\pnas}{Proc.~Nat.~Acad.~Sci.~USA}

\newcommand{\be}{{\bf e}}

\newcommand{\bv}{{\bf v}}
\newcommand{\bu}{{\bf u}}

\newcommand{\br}{{\bf r}}

\newcommand{\bk}{{\bf k}}

\newcommand{\bR}{{\bf R}}

\newcommand{\rcite}[1]{Ref.~\cite{#1}}
\newcommand{\rcites}[1]{Refs.~\cite{#1}}
\newcommand{\eq}[1]{Eq.~(\ref{#1})}
\newcommand{\eqs}[1]{Eqs.~(\ref{#1})}

\begin{document}

\title{Collective dynamics of chemically active particles trapped at a 
fluid interface}

\author{Alvaro Dom\'\i nguez}
\email{\texttt{dominguez@us.es}}
\affiliation{F\'\i sica Te\'orica, Universidad de Sevilla, Apdo.~1065, 
41080 Sevilla, Spain}

\author{P. Malgaretti}

\author{M. N. Popescu}

\author{S. Dietrich}

\affiliation{Max-Planck-Institut f\"ur Intelligente Systeme, Heisenbergstr.~3, 
70569 Stuttgart, Germany}

\affiliation{IV. Institut f\"ur Theoretische Physik, Universit\"{a}t Stuttgart,
Pfaffenwaldring 57, D-70569 Stuttgart, Germany}

\date{\today}

\begin{abstract}
Chemically active colloids generate changes in the chemical composition of their 
surrounding solution and thereby induce flows in the ambient fluid which 
affect their dynamical evolution. Here we study the many--body dynamics 
of a monolayer of active particles trapped at a fluid-fluid interface. To this 
end we consider a mean--field model which incorporates the direct pair interaction 
(including also the capillary interaction which is caused specifically by 
the interfacial trapping) as well as the effect of hydrodynamic interactions 
(including the Marangoni flow induced by the response of the interface to the 
chemical activity). The values of the relevant physical parameters for typical 
experimental realizations of such systems are estimated and various scenarios, 
which are predicted by our approach for the dynamics of the monolayer, are 
discussed. In particular, we show that the chemically--induced Marangoni flow can 
prevent the clustering instability driven by the capillary attraction. 
\end{abstract}

\maketitle

\textit{Introduction.}--- Chemically active, micron--sized particles are capable 
of self--induced motility by promoting a chemical reaction in the surrounding 
solution \cite{ebbens,SenRev}. Numerous experimental 
\cite{SenRev,ebbens,Golestanian2012,Fisher2014,Sanchez2015,Baraban2012,
Golestanian2014,brown14,Stocco2015} and theoretical 
\cite{Golestanian2005,Golestanian2007,Kapral2007,Julicher,Popescu2011EPL,
Kapral2013,Lowen2011,Michelin2015,Bechinger2014,Gompper2015,Stark2016,deGraaf2015} 
studies have been devoted to the motility mechanisms of such particles. 
Since the motion is caused by an intricate coupling between the chemical and 
hydrodynamic flow fields produced by these particles, this kind of colloids may 
exhibit a very complex behavior --- such as the emergence of surface--bounded 
steady--states \cite{Pine2013,Uspal2015a,Howse2015,Simmchen2016,Koplik2016},
enhanced motility \cite{MPD16}, rheotaxis \cite{Uspal2015b}, gravitaxis 
\cite{Ebbens2013,Stark2011} --- when they move near walls or are exposed to 
external flows or force fields.

The phenomenology becomes even richer if the walls, in addition to
their ``inert'' role as a provider of confinement, are themselves
responsive to, e.g., the chemical inhomogeneities induced by the
activity of the colloid \cite{Uspal2016}. A natural realization of
this scenario is a fluid--fluid interface, the surface tension of
which depends on the distribution of chemicals near the interfacial
region.  In this case, the spatially nonuniform distribution of
chemical components gives rise to Marangoni stresses\footnote{We remark that a similar motility
  mechanism can originate from thermally induced Marangoni flows if
  the particle is heated (see, e.g., \rcite{Wurger2014}).
  }
and the ensuing Marangoni flow in the ambient fluid leads either to 
self--propulsion along the interface for a particle trapped at the interface 
\cite{Lauga2011,Stone2014} or to an effective, long-ranged interaction of the 
particle with an interface located in its proximity \cite{Alvaro2016}.

Naturally the issue arises concerning the collective behavior of 
chemically active particles forming a monolayer at a responsive fluid--fluid interface.
The effective interaction between two particles, a distance $d$ apart, due to 
the advection by the induced Marangoni flow is equivalent to a long--ranged 
interparticle force decaying as $1/d$. The study in \rcite{MaSh14} has 
shown that on large scales this can give rise to a clustering instability similar 
to the ``chemotactic collapse'' exhibited by the Keller--Segel model 
\cite{Keller1970} if the chemical activity \textit{increases} the surface 
tension. However, in general, the particles exhibit also interactions which 
are not of hydrodynamic nature but determine the equilibrium states. (In 
the present study, for reasons of simplicity and for this kind of interactions we 
shall use the notion ``equilibrium-like''.) The capillary interaction between
particles trapped at a fluid interface is particularly notable 
\cite{KrNa00,OeDi08,Domi10}. In two respects this interaction differs strongly 
from, e.g., electrostatic double layer or van der Waals interactions: (i) it is a 
direct consequence of surface tension; (ii) it decays as $1/d$ over a wide range of 
significant interparticle separations of the colloids and induces a clustering 
instability formally analogous to a ``gravitational collapse'' \cite{DOD10,BDOD14} 
(and equivalent to the chemotactic collapse 
\cite{ChSi08}). These two features are shared with the Marangoni--induced dynamics, 
and a competition between both effects (i.e., capillarity and Marangoni advection) 
is conceivable. Furthermore, the particle motion driven by the ``equilibrium-like'' 
interactions also induces a flow --- formally indistinguishable from a 
Marangoni--stress driven one --- in the ambient fluid, which leads to ``anomalous 
collective diffusion'' \cite{BDGH14}.

Accordingly, in the present study we aim at understanding the collective
behavior exhibited by a monolayer of chemically active colloidal
particles trapped at an interface in the presence of \textit{both}
``equilibrium-like'' and hydrodynamic interactions. For this purpose,
we employ a theoretical framework developed previously in \rcites{DOD10,BDO15} for studying the collective dynamics of inert
colloids trapped at a fluid interface, and consistently incorporate
the simple model of chemical activity as discussed in \rcites{MaSh14,Alvaro2016}. We study the stability of a homogeneous
monolayer, with a special focus on the case that the
``equilibrium-like'' part of the interaction is dominated by the
monopolar capillary attraction. Since the capillary attraction is
exponentially screened beyond the capillary length $\lambda$ (which is of the
order of millimeter), the chemically induced Marangoni flow
eventually dominates at the largest scales.  However, on submillimeter
scales, the capillary forces (and, more specifically, the advection by
the ambient flow driven by them and leading to ``anomalous
diffusion'' \cite{BDGH14}) can dominate over the Marangoni flow. For typical
experimental conditions, we show that the Marangoni flow can prevent
the clustering driven by capillary attraction if the produced chemical species 
tends to \textit{decrease} the surface tension of the interface.

$\;$

\textit{Theoretical model.---} We consider a collection of particles
trapped at a flat fluid interface and forming a monolayer. The
dynamic evolution of the system is driven by Brownian diffusion;
external force fields; ``equilibrium-like'' particle interactions through
direct forces (e.g., hard--core repulsion, electrostatic forces, or,
specific to the presence of a fluid interface, capillary forces); and
interactions mediated by the surrounding ambient fluid (i.e., in the
form of hydrodynamic interactions).  The source of the latter is not
only the force (i.e., the ``hydrodynamic monopole'') acting on each particle,
but also the Marangoni stresses at the fluid interface induced by the
spatial distribution of a tensioactive chemical species ``$A$'' 
which is liberated (or absorbed) by the particles.

In order to describe the collective evolution of the monolayer we assume that 
the evolution of the areal particle density $\varrho(\br,t)$ occurs on time scales 
larger than any other process. (In order to keep the notation simple, we shall 
explicitly indicate the time dependence of $\varrho$ and other variables 
only when necessary.) Accordingly, one introduces the following fields 
(here, $\br=(x,y)$ denotes a point at the fluid interface, identified with the 
plane $z=0$): the velocity field $\bv(\br)$ of the monolayer; the (thermodynamic) 
force per particle $\mathbf{f}(\br)$ due to mutual interactions, external force 
fields, and Brownian diffusion; the number density $c(\br,z)$ of the chemical 
$A$ in bulk; the velocity field $\bu(\br,z)$ in the bulk of the ambient fluid 
above and below the fluid interface; and the inhomogeneous surface tension 
$\gamma(\br)$ of the fluid interface. (We note that all these fields are also 
(implicitly) time dependent through their dependence on $\varrho$.)

The density field obeys the continuity equation for the monolayer, expressing 
particle number conservation:
\begin{equation}
  \label{eq:rho}
  \frac{\partial \varrho}{\partial t} = - \nabla \cdot \left( \varrho \bv \right) ,
  \qquad
  \nabla := \left( \frac{\partial }{\partial x}, \frac{\partial }{\partial y} \right),
\end{equation}
with the velocity field $\bv(\br)$ in the overdamped regime of particle motion 
in the ambient flow given by
\begin{equation}
  \label{eq:v}
  \bv(\br) = \Gamma(\varrho) \mathbf{f}(\br) + \bu(\br,z=0) ,
\end{equation}
where $\Gamma(\varrho)$ is the particle mobility in the monolayer. This velocity 
field expresses the superposition of the drag by the force and the advection by 
the ambient flow. Under the assumption of local equilibrium in isothermal 
conditions, the thermodynamic force $\mathbf{f}(\br)$ (i.e., the negative gradient of 
the chemical potential) can be expressed as 
\begin{equation}
  \label{eq:f}
  \mathbf{f}(\br) = - \nabla \frac{\delta \mathcal{F}[\varrho]}{\delta \varrho(\br)}
\end{equation}
in terms of a free energy functional $\mathcal{F}[\varrho]$ for the monolayer. 
This functional accounts for the effect of Brownian motion, external force fields, 
and the ``equilibrium-like'' interactions between the particles. By 
construction this force has only an in-plane component; any non-vanishing component 
of the forces in the direction normal to the interface is exactly canceled by the 
constraining forces (usually the wetting forces) which impose the trapping at the 
interfacial plane $z=0$.

The ambient flow $\bu(\br,z)$ is given as the solution of the Stokes equation 
(describing incompressible flows at low Reynolds number) and it accounts 
both for the forces acting on the particles and for the Marangoni stresses 
at the fluid interface \cite{MaSh14,Alvaro2016,BDO15}:
\begin{equation}
  \label{eq:u}
  \bu(\br,z) = \int d^2\br' \; \left[ 
    \varrho(\br') \mathbf{f}(\br') + \nabla'\gamma(\br') 
  \right] \cdot \mathcal{O}(\br-\br'+z \be_z) ,
\end{equation}
with the Oseen tensor
\begin{equation}
  \label{eq:oseen}
 \mathcal{O}(\bR = \br+z\be_z) = \frac{1}{8\pi\eta_+ |\bR|} 
  \left[ \mathcal{I} + \frac{\bR \bR}{|\bR|^2} \right]\,,
\end{equation}
where $\mathcal{I}$ is the identity tensor and $\eta_+$ is the arithmetic mean 
of the dynamical viscosities of the upper and the lower fluid phases, 
respectively\footnote{We use the convention that adjacent 
vectors without the dot or the cross product sign denotes a dyadic 
(or tensorial) product.}. 
(Equation (\ref{eq:oseen}) is obtained from the general solution given 
in \rcite{JFD75} upon specializing to the case that the sources of the flow 
are localized within the plane $z=0$.)

Following the analysis in \rcite{Alvaro2016}, the dependence of the surface 
tension on the density of the tensioactive chemical at the interface is modeled as 
\begin{equation}
  \label{eq:gamma}
  \gamma(\br) = \gamma_0 - b_0 \left[ c(\br,z = 0) - c_0 \right],~~
  b_0 := \left. -\frac{d\gamma}{d c}\right|_{c = c_0},
\end{equation}
where $c_0$ and $\gamma_0$ are given reference values for the density of the 
species $A$ and the surface tension, respectively. Typically the coefficient 
$b_0$ is positive so that the interfacial tension is reduced by the presence 
of the chemical. 

The distribution of the chemical is determined as the stationary solution 
of the diffusion equation for the corresponding field $c(\br,z)$ with the 
particles acting as sources (or sinks) of the chemical species, assuming that the 
advection by the ensuing flows is negligible (i.e., the P{\'e}clet number of the 
species $A$ is considered to be very small \cite{Lauga2011,Stone2014,Alvaro2016}):
\begin{equation}
  \label{eq:c}
  D \left( \nabla^2 + \frac{\partial^2}{\partial z^2} \right) c 
  = - Q \delta (z) \varrho(\br) ,
\end{equation}
where $Q$ is the rate of production (if $Q>0$) or annihilation (if $Q<0$), 
respectively, of molecules of species $A$ per colloidal particle, and $D$ is the 
diffusion coefficient of the species $A$ in the fluids. For reasons of simplicity 
and in order to allow for an analytical derivation in closed form, 
rendering physically intuitive results, we take the diffusion coefficient as 
well as the solubility of the species to have the same value in both fluids. (Under 
these conditions, both the field $c$ and its derivatives are continuous at the 
interfacial plane \cite{Alvaro2016}. This simplifies the problem significantly 
because the presence of the interface does not influence the distribution of the 
chemical species.)

Finally, this mathematical model has to be complemented with appropriate 
boundary conditions. One usually considers (also for reasons of simplicity) 
an unbounded domain with appropriate boundary conditions at infinity, such as 
vanishing ambient flow and a fixed current or a fixed chemical potential for 
the chemical species $A$.

Equations (\ref{eq:rho}--\ref{eq:c}) form a closed system which determines 
the evolution of the particle density field $\varrho(\br,t)$. Before 
proceeding with the corresponding analysis, three remarks are in order.\newline
\textbf{(i)} The ambient flow $\bu(\br,z)$, entering \eq{eq:v} and being given
self--consistently in terms of the particle distribution, represents
the effect of the hydrodynamic interactions, i.e., how the ambient
flow driven by one particle affects the motion of the other ones.
There are two mechanisms leading to a hydrodynamic interaction, which according 
to \eq{eq:u} enter on equal footing: one contribution is due to the forces 
$\mathbf{f}(\br)$ acting on the particles, and one is due to the chemical activity 
of the particles and is represented by the Marangoni stresses $\nabla\gamma(\br)$. 
This flow is computed in the point--particle approximation: only the monopolar
(Stokeslet) term, given by the Oseen tensor, is retained, which describes a 
long--ranged hydrodynamic interaction (the Oseen tensor in \eq{eq:oseen} 
decays $\sim 1/|\mathbf{R}|$). This approximation can be viewed as the 
dilute limit of the model or, more generally, as a mean--field approximation 
in which the effect of the short--ranged correlations (in the form of
higher--order hydrodynamic multipoles) is incorporated --- at this level of 
description --- by means of an effective density--dependence of the 
rheological parameters of the monolayer as, e.g., the particle mobility $\Gamma$ 
\cite{Nozi87,Feld88}. Notwithstanding the common features just discussed, in 
the following we shall reserve the notion ``hydrodynamic interactions'' 
to those induced by the forces $\mathbf{f}(\br)$, as it is common use in the 
literature, and the effects by the Marangoni stresses will be referred to 
specifically.\newline
\textbf{(ii)} In \eq{eq:c} the particles are modeled as monopolar sources
of the tensioactive species $A$. This neglects, for reasons of simplicity, the 
detailed spatial structure of the production of $A$ on the surface of the 
particles, e.g, to which extent they are Janus particles. In particular, 
because of the assumed spherical symmetry (monopolar source), a single particle 
does not move by self--phoresis nor it is dragged in-plane by the 
Marangoni flow it induces. Therefore the influence of the particle activity on the 
dynamics shows up only as a collective effect via the hydrodynamic interactions and 
the Marangoni flows.\newline 
\textbf{(iii)} Equations (\ref{eq:rho}--\ref{eq:c}) encompass, as limiting 
cases, two situations addressed recently in the literature. If the particles are
not active (i.e., $Q=0$ in \eq{eq:c}), one recovers the model introduced in 
\rcite{BDGH14} for the emergence of anomalous diffusion due to the hydrodynamic 
interactions. If, however, the monolayer is modeled as a two-dimensional (2D)
ideal gas and the hydrodynamic interactions are neglected (i.e., if 
$\mathbf{f}=0$ in \eq{eq:u}), the model reduces to the one studied in 
\rcite{MaSh14}.

$\;$

\textit{Linear stability of the homogeneous state.}--- 
We restrict the discussion to the case in which the external force fields have 
vanishing in-plane components (e.g., gravity for a horizontal interface). 
Therefore they do not contribute to $\mathbf{f}(\br)$ and the whole spatial 
dependence of the free energy $\mathcal{F}$ in \eq{eq:f} enters only via
the density field. In this case, the model described by 
Eqs.~(\ref{eq:rho}--\ref{eq:c}) has a stationary 
solution given by the homogeneous distribution $\varrho(\br,t) = \varrho_0$. In the 
following we analyze the linear stability of this state with respect to small 
perturbations $\delta \varrho(\br,t) := \varrho(\br,t) - \varrho_0$.

By linearizing the governing equations we obtain the following system of equations 
which determines the evolution of the perturbation $\delta \varrho$:
\begin{subequations}
  \begin{equation}
    \label{eq:linrho}
    \frac{\partial\delta\varrho}{\partial t} 
    = - \varrho_0 
    \nabla \cdot \left[ \Gamma_0\delta\mathbf{f}(\br) 
      + \delta \bu(\br,z=0) \right] ,
  \end{equation}
where $\Gamma_0 := \Gamma(\varrho_0)$,
  \begin{equation}
    \label{eq:linf}
    \delta\mathbf{f}(\br)
    = - \nabla \int d^2\br' \; \left. 
      \frac{\delta^2 \mathcal{F}[\varrho]}{\delta \varrho(\br) \delta \varrho(\br')} 
    \right|_{\varrho=\varrho_0}
    \delta\varrho(\br') ,
  \end{equation}
and
\begin{widetext}
  \begin{equation}
    \label{eq:linu}
    \delta \bu(\br,z) = 
    \int d^2\br' \; \left[ \varrho_0 \delta\mathbf{f}(\br') 
      - b_0 \nabla' \delta c(\br',z=0) \right] \cdot
    \mathcal{O}(\br-\br'+z \be_z) ,
  \end{equation}
\end{widetext}
with $\delta c$ solving
  \begin{equation}
    \label{eq:linc}
    D \left( \nabla^2 + \frac{\partial^2}{\partial z^2} \right) \delta c 
    = - Q \delta (z) \delta\varrho(\br). 
  \end{equation}
\end{subequations}
Assuming that at infinity the perturbations vanish sufficiently fast, one can 
introduce the 2D Fourier transforms of the fields:
\begin{equation}
  \delta\varrho_\bk (t) := \int d^2\br \; \mathrm{e}^{-i\bk\cdot\br} 
  \delta\varrho(\br,t)
\end{equation}
and similarly for the others. With these \eq{eq:linrho} turns into
\begin{equation}
  \label{eq:fourierrho}
  \frac{\partial\delta\varrho_\bk}{\partial t} 
  = - i \bk \cdot \left[ \Gamma_0 \varrho_0 \delta\mathbf{f}_\bk
    + \varrho_0 \delta \bu_\bk(z=0) \right] .  
\end{equation}
The absence of in-plane contributions from external fields implies translational 
invariance and isotropy for the dynamics at the interface, which allows us to 
introduce a function $\hat{\mathcal{D}}_0(r)$ as
\begin{equation}
  \label{eq:hatD}
  \left. 
    \frac{\delta^2 \mathcal{F}[\varrho]}{\delta \varrho(\br) \delta \varrho(\br')} 
  \right|_{\varrho=\varrho_0} := 
  \frac{1}{\Gamma_0 \varrho_0}\hat{\mathcal{D}}_0(|\br-\br'|).
\end{equation}
Therefore the convolution on the right hand side of \eq{eq:linf} is given by 
\begin{equation}
  \label{eq:fourierf}
  \delta\mathbf{f}_\bk = 
  - i \bk \frac{\mathcal{D}_0(k)}{\Gamma_0 \varrho_0} \delta\varrho_\bk ,
\end{equation}
where
\begin{equation}
  \mathcal{D}_0(k) := \int d^2\br \; \mathrm{e}^{-i\bk\cdot\br} \,
  \hat{\mathcal{D}}_0(r)
\end{equation}
is the wavenumber--dependent coefficient of 2D collective diffusion of the 
monolayer in the absence of (long--ranged) hydrodynamic interactions (i.e., for 
$\delta\bu\equiv 0$ in \eq{eq:linrho}). The value $\mathcal{D}_0(k\to 0)$
is related to the isothermal compressibility of the monolayer in the equilibrium 
state \cite{dhon96}; for instance, for an ideal gas at temperature $T$ one has 
$\mathcal{F}_\mathrm{ideal}[\varrho] = k_B T \int d^2\br\; \varrho(\br) 
[ \ln [\Lambda^2 \varrho(\br)] - 1]$ (where $\Lambda$ and $k_B$ are de Broglie's 
thermal length and Boltzmann's constant, respectively) so that 
$\mathcal{D}_0(k)=\Gamma_0 k_B T$.

By using the known three-dimensional (3D) Fourier transform \cite{KiKa91} of 
the Oseen tensor one obtains its 2D Fourier transform as
\begin{eqnarray}
\label{eq:FT_Oseen}
&&  \int d^2\br \; \mathrm{e}^{-i\bk\cdot\br} \mathcal{O}(\br+z\be_z)
= \nonumber\\
 && \int_{-\infty}^{+\infty} \frac{dq}{2\pi} \; \mathrm{e}^{i q z}
  \frac{1}{\eta_+ (k^2+q^2)}\left[ 
    \mathcal{I} - \frac{(\bk + q\be_z) (\bk + q\be_z)}{k^2+q^2} 
  \right] = 
  \nonumber \\
  && \frac{\mathrm{e}^{-k|z|}}{4\eta_+ k} \left[ 2 \mathcal{I} 
    - (1-k|z|) \be_z \be_z - (1+k|z|) \frac{\bk \bk}{k^2} \right] ,
\end{eqnarray}
so that the convolution on the rhs of \eq{eq:linu} leads to
\begin{equation}
  i\bk\cdot\delta\bu_\bk(z=0) = \frac{i\bk}{4\eta_+ k} \cdot 
  \left[ \varrho_0 \delta\mathbf{f}_\bk - i \bk b_0 \delta c_\bk(z=0) \right] .
\end{equation}
Equation (\ref{eq:linc}) can be solved via the 3D Fourier transform of $\delta c$
from which one determines the 2D Fourier transform $\delta c_\bk(z=0)$
(analogously to the above calculation for the Oseen
tensor)\footnote{The divergences both in
      \eq{eq:FT_Oseen} and in \eq{eq:fourierc} as $k\to 0$
      reflect the slow spatial decay of the Green functions of
      the Stokes equation and the Poisson equation (\ref{eq:c}),
      respectively. They pose no mathematical difficulty because the Green
      functions are convoluted with the perturbation fields, which are taken 
      to vanish at infinity sufficiently fast.
}
\begin{equation}
  \label{eq:fourierc}
  \delta c_\bk(z=0) 
  = \int_{-\infty}^{+\infty} \frac{d q}{2\pi} \; 
  \frac{Q \, \delta\varrho_\bk}{D (k^2 + q^2)}
  = \frac{Q}{2 D k} \delta\varrho_\bk .
\end{equation}
By inserting Eqs.~(\ref{eq:fourierf}--\ref{eq:fourierc}) into \eq{eq:fourierrho}, 
one finally arrives at
\begin{equation}
  \label{eq:tau}
  \frac{\partial\delta\varrho_\bk}{\partial t} 
  = \frac{\delta \varrho_\bk}{\tau_k} ,
  \quad
   \frac{1}{\tau_k} = - k^2 \mathcal{D}_0(k) g(k)
  - \frac{\mathrm{sign}(Q b_0)}{T_\mathrm{surf}} ,
\end{equation}
where 
\begin{equation}
\label{eq:g_def}
g(k) := 1 + \frac{1}{L_\mathrm{hydro} k} , 
\end{equation}
with the length scale \cite{BDGH14}
\begin{equation}
  \label{eq:Lhydro}
  L_\mathrm{hydro} := \frac{4\eta_+\Gamma_0}{\varrho_0}
\end{equation}
and the time scale \cite{MaSh14}
\begin{equation}
  \label{eq:Tsurf}
  T_\mathrm{surf} := \frac{8\eta_+ D}{|Q b_0| \varrho_0} \,.
\end{equation}

Equation (\ref{eq:tau}) allows one to transparently identify the various 
driving mechanisms. The Marangoni stresses induce growth (if $Q b_0<0$) or decay 
(if $Q b_0>0$) of the perturbation at the same rate $1/T_\mathrm{surf}$ at all 
length scales, i.e., independent of $k$. Since typically $b_0>0$, the associated 
Marangoni flow leads to an effective repulsive interaction between the 
particles if they
are sources ($Q>0$) of the tensioactive agent. The diffusive dynamics
is encoded in $\mathcal{D}_0(k)$, while the factor $g(k)$ accounts for
the (long--ranged part of the) hydrodynamic interactions. Since
$g(k)>0$, the generic effect of these interactions is to reduce the
time scales associated with the diffusive evolution. This is at
variance with the usual effect of the short--ranged parts of the
hydrodynamic interactions, which tend to reduce the mobility
$\Gamma_0$ (at least for hard--spheres; the presence of direct short--ranged 
attractive interactions can favor an increase of the mobility
\cite{MLP10,LBP16}). Furthermore, as discussed in \rcite{BDGH14}, the factor 
$g(k)$ is responsible for the anomalous diffusion on length scales above 
$L_\mathrm{hydro}$: $g(k \to 0)$ diverges, which implies a 
rate $\sim k \mathcal{D}_0(0)$ for the evolution of the largest scales and thus 
amounts to a superdiffusive behavior. This divergence can be traced back to the 
dimensionality mismatch between the 2D dynamics of the confined colloid and the 3D 
hydrodynamic interactions mediated by the unconfined adjacent fluids.

$\;$

\textit{Discussion.---} First, we consider the case that the direct 
particle interactions are short--ranged, so that on large length scales 
$k^{-1}$ well above the relevant microscopic ones (e.g., the correlation length 
or the mean interparticle separation), one can approximate 
$\mathcal{D}_0(k)\approx \mathcal{D}_0(0) =:\mathcal{D}_\mathrm{short}$ (with 
the subscript ``short'' referring to the case of short--ranged interactions). 
Accordingly, \eq{eq:tau} states that for modes with 
sufficiently small values of $k$ the dynamics is always dominated by the effect 
of the Marangoni flow. Therefore, even if the homogeneous state is thermodynamically 
stable ($\mathcal{D}_\mathrm{short}>0$), for $Q b_0<0$ a clustering instability 
is predicted. This instability is driven by the Marangoni flow and occurs for 
modes with sufficiently small wavenumbers,
\begin{equation}
  \label{eq:kcrit}
  k < k_c:= \frac{1}{2 L_\mathrm{hydro}} \left[ 
    \sqrt{1 + \left(\frac{2 L_\mathrm{hydro}}{L_\mathrm{surf}}\right)^2} - 1
  \right] ,
\end{equation}
where $k_c$ is the positive solution of $1/\tau_k = 0$ in \eq{eq:tau}, 
and the length scale $L_\mathrm{surf}$ is given by
\begin{equation}
  \label{eq:Lsurf}
  L_\mathrm{surf} := \sqrt{\mathcal{D}_\mathrm{short} T_\mathrm{surf}} \,.
\end{equation}
If $L_\mathrm{hydro}\gg L_\mathrm{surf}$, this expression reduces to
$k_c \approx 1/L_\mathrm{surf}$ and the regime (discussed recently in
\rcite{MaSh14}) in which the hydrodynamic interactions are neglected 
(i.e., $g(k)\approx 1$ in \eq{eq:tau}) is recovered. In the opposite limit, 
$L_\mathrm{hydro}\ll L_\mathrm{surf}$, this critical wavenumber is reduced 
by a large factor down to $k_c \approx L_\mathrm{hydro}/L_\mathrm{surf}^2$ 
because the diffusion (enhanced into super-diffusion by the hydrodynamic 
interactions) is more effective in stabilizing the homogeneous state.

We now consider the case that the particles interact among each other 
through capillary forces due to deformations of the fluid interface. This 
choice of the interaction is particularly interesting because, like the 
Marangoni stresses, these forces are 
specific to the presence of a fluid interface and have a strength related to the 
value of the surface tension. The capillary interaction can be modeled 
analogously to direct electrostatic interactions in the limit that the 
deviation of the 
interface from the flat state is small \cite{DOD08}. An experimentally relevant 
example is the one in which the interfacial deformation arises because 
each particle
experiences a vertical force $F \be_z$ due to an 
external  agent such as buoyancy or an external electric field normal to 
the interface \cite{ASJN08}. (As remarked below \eq{eq:f}, 
only in-plane interactions contribute to the force $\mathbf{f}$; thus 
$\mathbf{F}$, which is normal to the interface, does not affect $\textbf{f}$.)  
In this case, and within the point--particle approximation, the particles can 
be described as so-called ``capillary monopoles''. The associated interaction 
is attractive and has a range of the order of the capillary length\footnote{The 
capillary length is given by $\lambda_0 = \sqrt{\gamma_0/(g\, \Delta\rho)}$, where 
$\gamma_0$ is the surface tension, $g$ is the acceleration of gravity, and $\Delta\rho$ 
is the mass density contrast between the coexisting fluids.} 
$\lambda_0$. For typical fluid interfaces this length is of the order of 
millimeter. 
Therefore, for colloidal--sized particles this capillary monopolar interaction can be successfully addressed within a mean--field approximation, leading to \cite{DOD10}
\begin{equation}
  \label{eq:monopoleD}
  \mathcal{D}_0(k) = - \frac{1}{T_{\mathrm{Jeans}}}\,\,
  \frac{\lambda_0^2}{1+(\lambda_0 k)^2} ,
\end{equation}
with the characteristic time (Jeans time) 
\begin{equation}
  \label{eq:Tjeans}
  T_{\mathrm{Jeans}} := \frac{\gamma_0}{\Gamma_0 F^2 \varrho_0} .
\end{equation}
(This result follows by assuming a spatially constant surface tension 
$\gamma_0$. Within the linear approximation of small perturbations
$\delta \rho$ considered here, the correction to the capillary interaction 
due to a spatially varying surface tension does not contribute.)
The fact that the range $\lambda_0$ is much larger than the
interparticle separation sets this interaction apart from the other
ones mentioned above: there is a whole range of wavenumbers $k$ with 
$k \to 0$ but $\lambda_0 k \gg 1$ for which the continuum description we have
introduced is valid and yet the approximation $\mathcal{D}_0(k) \sim
k^{-2}$ (\eq{eq:monopoleD}) holds. In this intermediate range of scales 
this formal divergence is a signature of the long--ranged nature of the monopolar 
capillary attraction  (which is formally analogous to 2D Newtonian
gravity \cite{BDOD14}). In \eq{eq:tau} it leads  to a $k$--independent
contribution like the one by the Marangoni flow, thus emphasizing the
importance of studying the possible competition between these two
effects. We remark that \eq{eq:monopoleD} describes an instability
($\mathcal{D}_0(k)<0$, the so-called ``capillary collapse'') because
for reasons of simplicity we have neglected the stabilizing effect of
Brownian diffusion and of possible short--ranged repulsions. (In 
\eq{eq:monopoleD} this
  would appear as an additive term $\mathcal{D}_\mathrm{short}$ and it 
  would become 
effective on length scales
  below the Jeans length $L_{\mathrm{Jeans}} :=
  \sqrt{\mathcal{D}_\mathrm{short} T_{\mathrm{Jeans}}}$ \cite{DOD10}). 

\begin{figure*}[!htb]
\includegraphics[width=\textwidth]{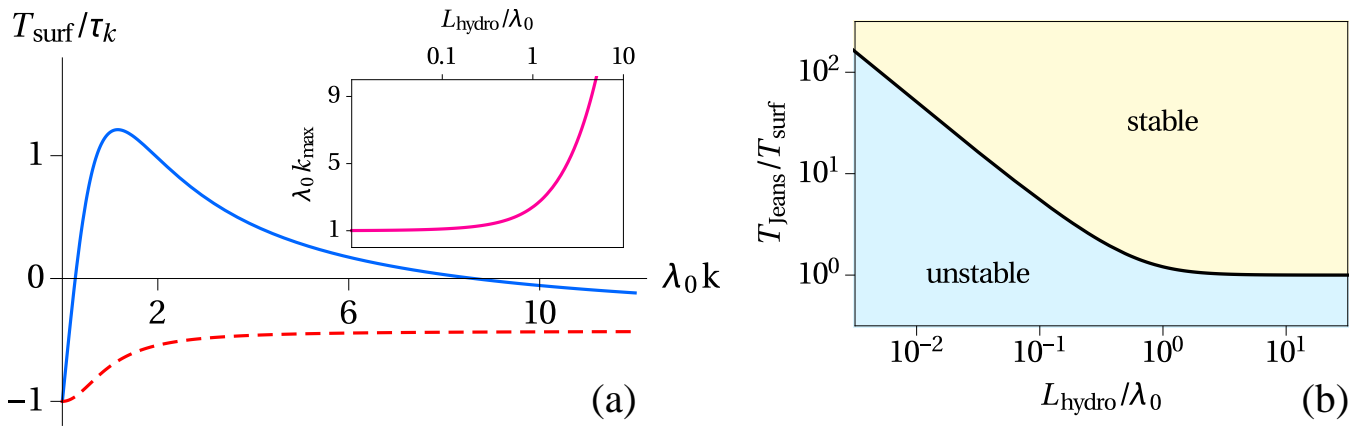}
\caption{(a) The growth rate $\tau_k^{-1}$ (solid blue line,
\eq{eq:taucap}) for $Q b_0>0$ and the particular parameter choices 
$L_\mathrm{hydro}/\lambda_0=0.15$ and $T_{\mathrm{Jeans}}/T_\mathrm{surf} = 
1.75$. One observes a range of modes with $\lambda_0 k \sim 1$ which are unstable 
($\tau_k^{-1} > 0$). For comparison, the dashed red line shows $\tau_k^{-1}$ 
for $L_\mathrm{hydro}\to\infty$; in this case all modes are stable. 
The inset shows the value $k_\mathrm{max}$ of the wavenumber at which the growth 
rate is maximum (see \eq{eq:kmax}). (b) Stability diagram in the 
parameter space spanned by the ratios $L_\textrm{hydro}/\lambda_0$ and 
$T_{\textrm{Jeans}}/T_\textrm{surf}$. The separatrix is determined 
by $\tau_{k_\mathrm{max}}^{-1}=0$. Accordingly, ``stable'' (``unstable'') 
means that all modes (some modes) are stable (unstable) within the present 
linear stability analysis.
}
\label{fig:monopole}
\end{figure*}
As discussed above, the sign of the product $Q b_0$ determines whether the effective 
interactions due to the Marangoni flows are attractive or repulsive. First, we 
consider the case $Q b_0 > 0$, so that the repulsive effect of the Marangoni flow could 
actually counterbalance the capillary instability. In this case, the combination of
\eqs{eq:tau} and (\ref{eq:monopoleD}) gives the growth rate
  \begin{equation}
    \label{eq:taucap}
    \frac{1}{\tau_k} = \frac{1}{T_{\mathrm{Jeans}}}\,\,
    \frac{(\lambda_0 k)^2}{1+(\lambda_0 k)^2} \;
    \left[ 1 + \frac{1}{L_\mathrm{hydro} k} \right]
    - \frac{1}{T_\mathrm{surf}} .
  \end{equation}
Thus, the relevant physical dimensionless parameters are the ratio 
$L_\textrm{hydro}/\lambda_0$, which controls the importance of the 
anomalous--diffusion effect on the capillary--driven dynamics, and the ratio 
$T_\textrm{Jeans}/T_\textrm{surf}$, which measures the relative strength of the 
capillary forces and the Marangoni flows. The growth rate $\tau_k^{-1}$ is 
illustrated in Fig.~\ref{fig:monopole}(a), which allows one to 
straightforwardly infer the possible scenarios.

On scales $k \ll \lambda_0^{-1}$, the capillary interaction is
screened and the behavior thus corresponds to the case discussed above
with $\mathcal{D}_0(k) \approx
\mathcal{D}_0(0)=-\lambda_0^2/T_{\mathrm{Jeans}}$ finite, i.e., the
stabilizing effect of the Marangoni flow dominates. However, on scales
$k \gtrsim \lambda_0^{-1}$ the system can be destabilized by the
capillary forces; this occurs if the maximum of the curve in
Fig. \ref{fig:monopole}(a) lies above zero. The wavenumber
$k_\mathrm{max}$ of the position of the maximum is given as
  \begin{equation}
    \label{eq:kmax}
    \lambda_0 k_\mathrm{max} = \frac{L_\mathrm{hydro}}{\lambda_0} + 
    \sqrt{ 1 + \left( \frac{L_\mathrm{hydro}}{\lambda_0} \right)^2 } ,
  \end{equation}
(see the inset in Fig.~\ref{fig:monopole}(a)). The boundary of stability in 
the parameter space is given by the condition $\tau_{k_\mathrm{max}}^{-1} = 0$, 
see Fig.~\ref{fig:monopole}(b). When $L_\mathrm{hydro} \ll \lambda_0$, 
which is the most 
natural case for a colloidal monolayer (see, c.f., \eq{eq:Lhydronum}), 
the effect of the long--ranged hydrodynamic interactions on the capillary--driven 
dynamics is important.  In this limit one has $\lambda_0 k_\mathrm{max} \approx 1$, 
and the computation of $\tau_{k_\mathrm{max}}$ shows that the homogeneous state is 
stabilized by the Marangoni flow against capillary collapse if
\begin{equation}
  \label{eq:stability}
  \frac{T_{\mathrm{Jeans}}}{T_\mathrm{surf}} >
  \frac{\lambda_0}{2 L_\mathrm{hydro}} \; (\gg 1) .
\end{equation} 
This inequality follows from requiring that $\tau_{k_\mathrm{max}}$ is 
negative. On the other hand, 
if this inequality does not hold, an intermediate range 
of length scales are unstable, with the modes $k \approx \lambda_0^{-1}$ growing the 
fastest. Although this is reminiscent of the early stages of spinodal decomposition 
during phase separation, further work is required in order to assess to what extent 
the nonlinear dynamical evolution is comparable with a coarsening scenario. 

The specific signature brought about by the hydrodynamic interactions can be 
identified by comparing their influence with the predictions in the opposite limit 
$L_\mathrm{hydro} \gg \lambda_0$, so that for the wavenumbers 
$k\gtrsim\lambda_0^{-1}$, for which the capillary instability is relevant, 
one can neglect the effect of the hydrodynamic interactions (by setting 
$g(k) \to 1$ in \eq{eq:tau}). The stability diagram in Fig.~\ref{fig:monopole} 
shows that in this latter case the conditions, under which 
stabilization of the homogeneous distribution by the Marangoni flow occurs, 
are less stringent: $T_{\mathrm{Jeans}}/T_\mathrm{surf} > 1$ is sufficient. 
If a mode is unstable, there is usually a broad range of wavenumbers, $1 
\lesssim \lambda_0 k \lesssim \lambda_0 k_\mathrm{max} 
\approx 2 L_\mathrm{hydro}/\lambda_0$, for which 
the corresponding modes all grow roughly at the same rate (as in the case of the 
purely capillary instability \cite{DOD10}).

We finally succinctly remark on the case $Q b_0 <0$: both
  capillarity and Marangoni flows induce an interparticle attraction
  and thus destabilize the homogeneous state. The two regimes, in which either one 
or the 
  other mechanism is dominant, can be visualized from a diagram which has an 
  appearance similar to Fig.~\ref{fig:monopole}(b), with the alternative 
reading that it
  depicts the parameter regions where clustering is driven
  predominantly either by capillary attraction (in the region
  labeled ``unstable'', as before) or by the Marangoni flows (in the
  region labeled ``stable'').

$\;$

\textit{Numerical estimates.---} In order to complement the
theoretical analysis, here we discuss the expected values of
the parameters $T_{\mathrm{Jeans}}/T_\mathrm{surf}$ and
  $L_\mathrm{hydro}/\lambda_0$ for typical experimental configurations
at room temperature. A particle is characterized by its
  radius $R$, the chemical activity $Q$, and the capillary monopole
  $F$. We introduce the dimensionless parameter $q$, which compares the 
strength of the Marangoni flow with the Brownian
motion \cite{Alvaro2016}, and the dimensionless parameter
$W_\mathrm{cap}$, which compares the strength of the inter-particle potential
due to capillary attraction  with the thermal energy:
\begin{equation}
\label{eq:def_q}
  q_R : = \dfrac{3 \,Q b_0 \,R}{32 \,D \,k_B T} \,,
  \qquad W_\mathrm{cap} := 
  \dfrac{F^2}{2 \pi \gamma_0 k_B T} \,.
\end{equation}
We also invoke the Stokes mobility for a single particle:
  \begin{equation}
    \label{eq:Gamma}
    \Gamma_0 = \frac{1}{6\pi\eta_+ R} \,.
  \end{equation}
  (Although we are dealing with monolayers, it is known that for colloids in 
bulk solution \eq{eq:Gamma} provides a reasonable, order-of-magnitude estimate of 
the mobility, which depends on the density and the interactions 
\cite{MLP10,LBP16}.) From \eqs{eq:Tsurf} and (\ref{eq:Tjeans}) one obtains the ratio
  \begin{equation}
    \frac{T_{\mathrm{Jeans}}}{T_\mathrm{surf}} 
    = \frac{4 \,|q_R|}{W_\mathrm{cap}}\,.
  \end{equation}
  For colloidal particles at an air--water or an oil--water
  interface ($\gamma_0 \sim 0.05 \; \mathrm{N}/\mathrm{m}$), buoyancy
  effects lead to an estimate for the capillary force in the order of 
$W_\mathrm{cap} \sim
  10^{-6} \times (R/\mathrm{\mu m})^6$ \cite{DOD10}, so that
  \begin{equation}
    \frac{T_{\mathrm{Jeans}}}{T_\mathrm{surf}} 
    \sim 4\times 10^6 \times |q_1|
    \left(\frac{R}{\mu\mathrm{m}}\right)^{-5} \,, 
  \end{equation}
  where $q_1: = q_R(R = 1~\mathrm{\mu m})$ (see \eq{eq:def_q})). The sensitive 
  dependence of the capillary monopole on the 
  particle size is transferred to this ratio.
The hydrodynamic length scale $L_\mathrm{hydro}$ can be estimated from
\eqs{eq:Lhydro} and (\ref{eq:Gamma}) as
  \begin{equation}
    \label{eq:Lhydronum}
    \frac{L_\mathrm{hydro}}{\lambda_0} = \frac{2}{3 \phi} \frac{R}{\lambda_0},
  \end{equation}
  where $\phi = \pi R^2 \varrho_0$ is the 2D packing fraction of the
  monolayer. For a not too dilute monolayer of colloidal
    particles, \eq{eq:Lhydronum} predicts that $L_\mathrm{hydro}/\lambda_0$ 
    is very small.
    
\begin{figure}
  \includegraphics[width=0.45\textwidth]{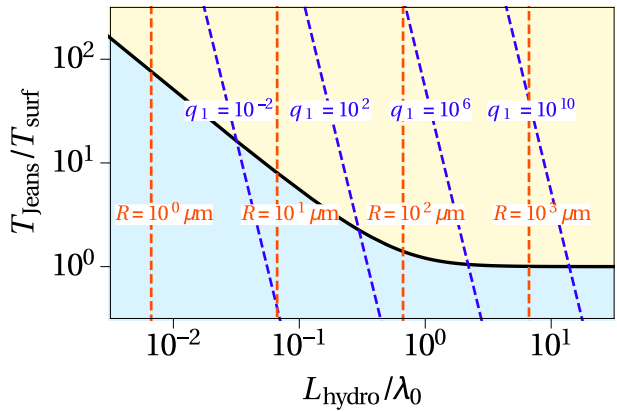}
  \caption{Plot of the iso-$R$ and iso-$q_1$ lines in the stability
    diagram for a monolayer with packing fraction $\phi=0.1$ at an
    interface with capillary length $\lambda_0=1\;\mathrm{mm}$.}
\label{fig:numerical}
\end{figure}
  Figure~\ref{fig:numerical} shows the curves of constant $R$, as well as those of 
  constant $q_1$, superimposed on the stability diagram
    for a not too dilute monolayer ($\phi=0.1$) and with the
    order-of-magnitude choice $\lambda_0=1\;\mathrm{mm}$. One observes
    that an increase in the chemical activity (i.e., the value of
    $q_1$) at constant $R$ or a reduction in $R$ at constant chemical
    activity promote stability of the homogeneous state.  As an
    illustration, the active particles described in \rcite{PKOS04}
    have $Q/(4\pi R^2) \sim 10^{-3}~\mathrm{mol}/(\mathrm{s}\times
    \mathrm{m}^2)$ (corresponding to platinum--covered particles
    catalyzing the decomposition of peroxide into water and oxygen, which 
    is weakly tensioactive). This leads
    to a small value $|q_1| \sim 10^{-2}$ at an air--water interface.
    However, switching to a liquid--liquid interface (for which the
    diffusivity $D$ of oxygen is strongly reduced), enhances the influence of
    the chemical activity up to $|q_1| \sim 10^2$ \cite{Alvaro2016} and
    the Marangoni flows can stabilize the homogeneous monolayer even
    for rather large particles.

  As a final remark, we notice that Brownian diffusion sets a
    lower limit on the particle size for the relevance of the
    capillary attraction: since $\mathcal{D}_\mathrm{short} = \Gamma_0
    k_B T$, the Jeans length is $L_{\mathrm{Jeans}} = \sqrt{\Gamma_0 k_B
      T \; T_{\mathrm{Jeans}}} = R/\sqrt{2\phi W_\mathrm{cap}}$, which
    becomes larger than the capillary length
    $\lambda_0=1\;\mathrm{mm}$ for radii below $1\;\mu\mathrm{m}$.

$\;$

\textit{Conclusions.---} We have presented a continuum description
which extends previous studies \cite{DOD10,BDGH14,MaSh14} to the
case of the dynamics of a monolayer of chemically active colloidal
particles trapped at a fluid interface.  The model accounts for capillary
and hydrodynamic interactions, the latter including the flow induced
\textit{both} by the particle motion and by the response of the interface
to the chemical activity (i.e., Marangoni flows). The model is a
mean--field theory, which is valid for the monopolar
contribution to the capillary and hydrodynamic interactions. Our
analysis of the linear stability of the homogeneous state revealed a
novel stabilization mechanism against clustering driven by
  capillary attraction, which is due to the induced Marangoni flows
(giving rise to repulsive effective interactions). Therefore this 
mechanism is a feature specific
to the ``chemically active'' nature of the particles and is operational at 
all length-scales, provided the ratio
$T_{\mathrm{Jeans}}/T_\mathrm{surf}$ (Eqs.~(\ref{eq:Tjeans}) and 
(\ref{eq:Tsurf})) of characteristic times is sufficiently large (see 
Fig.~\ref{fig:monopole}(b)). Our numerical estimates indicate that 
for typical experimental setups this can be the case (see 
Fig.~\ref{fig:numerical}). If this ratio is too small, the perturbations at the 
long and the very short wavelengths are stabilized by the Marangoni flows, 
but  an intermediate range of length
scales are unstable with the modes $k \approx \lambda_0^{-1}$ growing
the fastest (see Fig.~\ref{fig:monopole}(a)). These findings lend themselves to be 
explored further by employing numerical studies of the full dynamics.

\textit{Acknowledgements.---} A.D.~acknowledges support by
    the Spanish Government through Grant FIS2014-53808-P (partially
    financed by FEDER funds). Support from the COST Action MP 1305
  ‘Flowing matter’ is gratefully acknowledged by the authors.

\end{document}